\pdfoutput=1
\PassOptionsToPackage{pdfpagelabels=false}{hyperref}
\documentclass[fleqn,usenatbib,usedcolumn]{mnras}
\usepackage[british]{babel}             
\usepackage{txfonts}
\let\la=\lesssim 
\usepackage[T1]{fontenc}
\usepackage{ae,aecompl}

\usepackage{graphicx}	

\newcommand{\deltamagX}{\ensuremath{\Delta\textrm{F}_{\textrm{X}}}}
\newcommand{\deltamagB}{\ensuremath{\Delta\textrm{F}_{\textrm{B}}}}
\newcommand{\deltamagV}{\ensuremath{\Delta\textrm{F}_{\textrm{V}}}}
\newcommand{\deltamagR}{\ensuremath{\Delta\textrm{F}_{\textrm{R}}}}
\newcommand{\deltamagIR}{\ensuremath{\Delta\textrm{F}_{\textrm{IR}}}}

\title[GALAH Survey: Relative throughputs of 2dF/HERMES]{The GALAH Survey: Relative throughputs of the 2dF fibre positioner and the HERMES spectrograph from stellar targets}
\author[J. D. Simpson et al.]{Jeffrey D. Simpson$^{1}$\thanks{Email: \texttt{jeffrey.simpson@aao.gov.au}}, G. M. De Silva$^{1,2}$, J. Bland-Hawthorn$^2$, K. C. Freeman$^3$,
\newauthor{S. L. Martell$^4$, Katharine J. Schlesinger$^3$, Sanjib Sharma$^2$, D. B. Zucker$^{5,6}$, T. Zwitter$^7$,}
\newauthor{J. Kos$^2$, Borja Anguiano$^5$, David M. Nataf$^{3}$, Warren Reid$^{5,8}$, Robert A. Wittenmyer$^{4}$}\\
$^{1}$Australian Astronomical Observatory, North Ryde, NSW 2113, Australia\\
$^{2}$Sydney Institute for Astronomy, School of Physics, A28, The University of Sydney, NSW 2006, Australia\\
$^{3}$Research School of Astronomy and Astrophysics, Australian National University, Canberra, ACT 2611, Australia\\
$^4$School of Physics, University of New South Wales, Sydney, NSW 2052, Australia\\
$^5$Department of Physics and Astronomy, Macquarie University, Sydney, NSW 2109, Australia\\
$^6$Research Centre in Astronomy, Astrophysics \& Astrophotonics, Macquarie University, Sydney, NSW 2109, Australia\\
$^7$Faculty of Mathematics and Physics, University of Ljubljana, Jadranska 19, 1000 Ljubljana, Slovenia\\
$^8$Western Sydney University, Locked Bag 1797, Penrith South DC, NSW 1797, Australia
}

\date{Accepted 2016 March 29. Received 2016 March 20; in original form 2016 January 21}

\pubyear{2016}
\volume{{\rm in press}}

\begin{document}
\label{firstpage}
\pagerange{\pageref{firstpage}--\pageref{lastpage}}
\maketitle

\begin{abstract}
We present an analysis of the relative throughputs of the 3.9-metre Anglo-Australian Telescope's 2dF/HERMES system, based upon spectra acquired during the first two years of the GALAH survey. Averaged spectral fluxes of stars were compared to their photometry to determine the relative throughputs of fibres for a range of fibre position and atmospheric conditions. We find that overall the throughputs of the 771 usable fibres have been stable over the first two years of its operation. About 2.5 per cent of fibres have throughputs much lower than the average. There are also a number of yet unexplained variations between the HERMES bandpasses, and mechanically \& optically linked fibre groups known as retractors or slitlets related to regions of the focal plane. These findings do not impact the science that HERMES will produce.

\end{abstract}




\section{Introduction}\label{sec:intro}
Large spectroscopic surveys have changed our view of the Universe. Their success often relies on large field-of-view, highly-multiplexed, fibre-fed instruments that allow for the simultaneous observations of hundreds (and in some cases thousands) of targets. For this reason, wide-field, multi-object spectrographs are the main workhorse instruments at many leading observatories around the world.

On the 3.9-metre Anglo-Australian Telescope (AAT), the Two-degree Field fibre positioner (2dF) continues a strong tradition of innovation in multi-fibre spectroscopy at the Australian Astronomical Observatory (AAO). 2dF is a top-end mounted instrument that (in its current iteration) can position up to 392 optical fibres across a 2\degr\ field of view for multi-object spectroscopic observations \cite[for a detailed overview of 2dF in its original iteration, see][]{Lewis2002}. When originally commissioned for the 2dF Galaxy Redshift Survey \citep{Colless2001}, it had spectrographs mounted on the top end ring of the telescope, as at the time optical fibres were limited in their efficiencies over large distances, necessitating the spectrograph be as close as possible to minimize light losses (in this case, the fibres were 7.5 metres long).

With improvements in optical fibres, in 2006 the AAO replaced the 2dF-mounted spectrographs with a new spectrograph known as AAOmega \citep{Sharp2006}, which is located in a thermally isolated room with light fed to it from 2dF via a 38-metre long cable. Following the success of AAOmega in achieving good quality multi-object, fibre-fed spectra via the longer cable, a second 2dF-fed spectrograph was built: the High Efficiency and Resolution Multi-Element Spectrograph (HERMES), which achieved first light in 2013. Like AAOmega, HERMES is situated in an atmospherically stable room, again requiring a long fibre cable, in this case 50 metres.

A key advantage of fibres for multi-object spectrographs is the ability to reformat the wide field of view into a fixed format on the detector. However as each fibre needs to be individually handled in order to be placed correctly on the field of view, the fibres undergo a variety of stresses. There is currently little in the literature discussing the relative fibre transmission and the long-term stability of such systems, but it is critical that we understand the fibre properties for the future success of large multi-object spectrographs.

Because the fibres are configured as part of complex dynamical systems, wide-field multi-object spectrograph systems are notoriously difficult to flux calibrate. There are many free parameters, e.g., field position, dispersion correction, tension in fibres, absolute throughput efficiency of the optics and the telescopes, fibre age. In this paper we are only considering the relative transmission of the fibres.

The relative transmission and temporal degradation of fibres in the AAOmega cable were investigated by \cite{Sharp2013} using datasets provided by two long-term galaxy redshift surveys: WiggleZ \citep{Drinkwater2010} and GAMA \citep{Driver2011}. In both surveys a few fibres per field were allocated to Galactic field stars for calibration purposes. \cite{Sharp2013} used two different analysis methods: comparing the spectral brightness of the calibration stars to their previously determined photometry (the method used in our work); and using the bright OH sky emission lines, which are visible in the spectra from all the fibres.

In this paper we analyse the HERMES fibre cable, in a similar way as \cite{Sharp2013} did for the original AAOmega fibre cable, using observations acquired by the Galactic Archaeology with HERMES (GALAH) survey \citep{Silva2015}. GALAH, the primary science driver of HERMES and its largest user, is conducting a Milky Way stellar survey to reconstruct the history of our Galaxy's formation using precise multi-element abundances of one million stars.

\begin{figure}
	\includegraphics[width=\columnwidth]{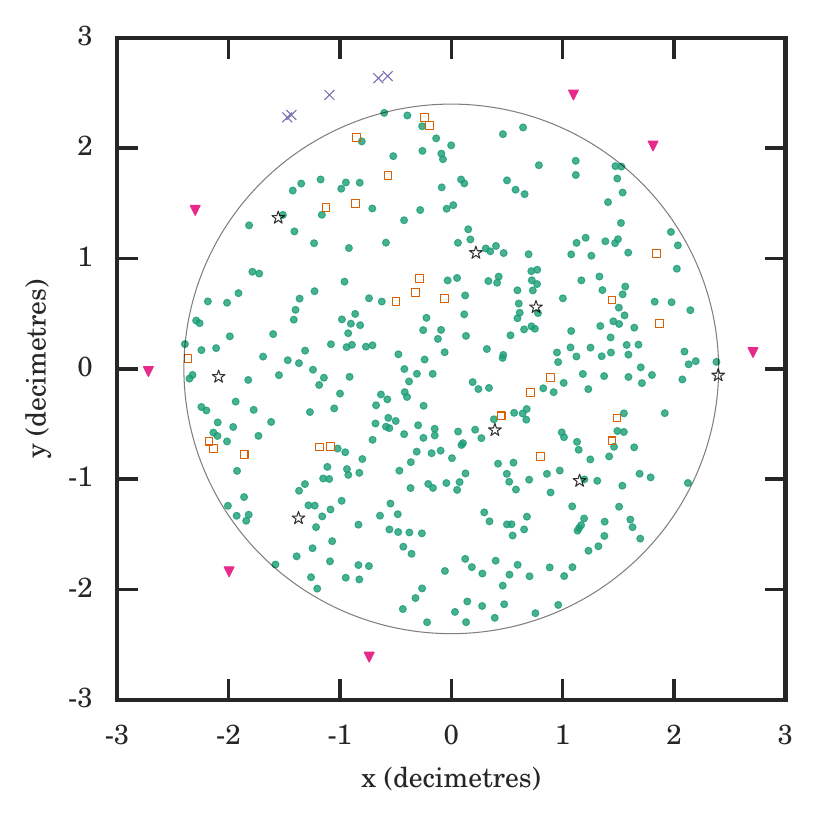}
    \caption{An example of the fibre arrangement on a typical GALAH field on a 2dF plate. Dots in green show the position of the fibres allocated to stellar targets, the open orange squares show the position of the fibres allocated to measuring the sky brightness (i.e., they are aimed at nominally empty regions of sky), and the star symbols for the special fiducial fibres which are allocated to bright ($V\sim11$) stellar targets and are used for auto-guiding the telescope. The black circle marks the edge of the on-sky region. Those fibres outside of the on-sky circle are parked, either because the allocation software could not find a solution in which to use them (purple crosses), or they are broken or unavailable for use with HERMES (pink triangles). The declination of the stars increases from left-to-right and right ascension increases from bottom-to-top.}
    \label{fig:typical_galah_field}
\end{figure}

For details on the observational strategy and survey design of GALAH see Martell et al. (2016, in prep). Briefly, an input catalogue was created from the union of the AAVSO Photometric All-Sky Survey (APASS) \citep{Henden2012,Henden2014,Munari2014}, the Two Micron All-Sky Survey (2MASS) \citep{Skrutskie2006}, and the Fourth US Naval Observatory CCD Astrograph Catalog (UCAC4) \citep{Zacharias2013} of all stars with V magnitudes between 12 and 14. This was then cut-down to regions of the sky where the following requirements were met: $-80\degr < \textrm{Dec} <+10\degr$ (visible from the AAT); stellar density on the sky of at least 400 GALAH target stars per 2\degr\ field (in order to fully populate the fibre configuration); Galactic latitudes $b>|10\degr|$ (so as not to be too close to the Galactic plane, where the stellar density on the sky is too high). This was sub-divided into about 6000 $2\degr$-fields of 400 stars each.

As GALAH is observing Galactic disk stars, the field configurations are typically uniformly random in the distribution across the focal plane (see Figure \ref{fig:typical_galah_field} for the plate configuration of a typical GALAH field). The magnitude and the colour distribution of stars will also be about the same for all the fields. Obviously these assumptions of sky distribution and colour properties break down in particular regions of the sky, such as when a bright cluster is within a field region, but overall there is good homogeneity across the survey fields.

This homogeneity makes the GALAH survey dataset an excellent choice for analysing the throughputs of the 2dF fibres and HERMES itself. Additionally HERMES is designed with four non-contiguous bandpasses, allowing us to make four semi-independent measures of the throughput at different wavelengths.

In this paper we will discuss the pertinent details of the spectrograph and fibre cable design (Section \ref{sec:design}); the reduction of the raw spectra (Section \ref{sec:datareduction}); the method by which the throughputs of the fibres were determined (Section \ref{sec:throughputs}); and the results for fibre-to-fibre and retractor-to-retractor variations in throughput (Section \ref{sec:results}). Finally we have a brief discussion about how upgrades to HERMES and the fibre cable could affect these throughputs in the future (Section \ref{sec:discussion}).

\section{Spectrograph and fibre cable design}\label{sec:design}

\begin{table}
\caption{Wavelength ranges of the HERMES's cameras, and the photometric equivalents that were used for each camera in this work. They are visually represented in Figure \ref{fig:bandpass_comparison}.}
  \label{table:wavelength_comparison}
  \begin{tabular}{lcc}
  \hline
   Camera & HERMES wavelength range & Photometric equivalent \\
    \hline
   Blue & 4715 -- 4900 \AA & B: 3613 -- 5113 \AA \\
   Green & 5649 -- 5873 \AA & V: 4796 -- 6455 \AA \\
   Red & 6478 -- 6737 \AA & r: 5415 -- 6989 \AA\ \\
   Infrared & 7585 -- 7887 \AA & i: 6689 -- 8389 \AA\ \\ 
   \hline  
  \end{tabular} 
\end{table}

\begin{figure}
	\includegraphics[width=\columnwidth]{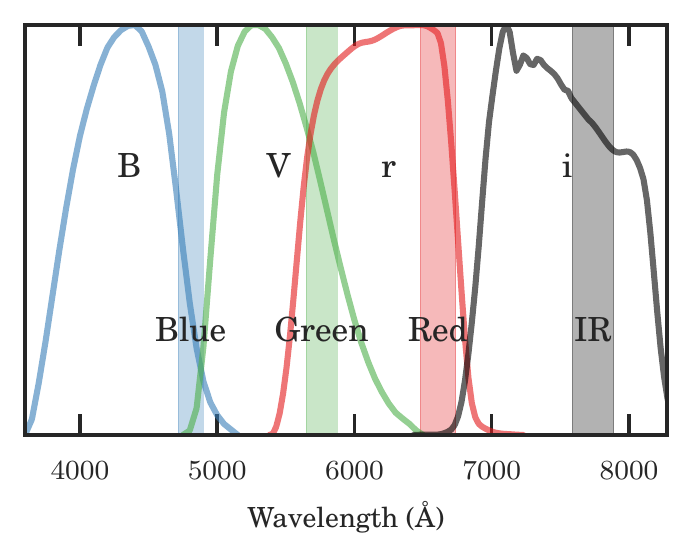}
    \caption{Comparison of the bandpasses of HERMES's cameras (solid bars), and the bandpasses of the photometry that was used for comparison with each camera. The wavelength ranges are also numerically given in Table \ref{table:wavelength_comparison}.}
    \label{fig:bandpass_comparison}
\end{figure}

\begin{figure}
	\includegraphics[width=\columnwidth]{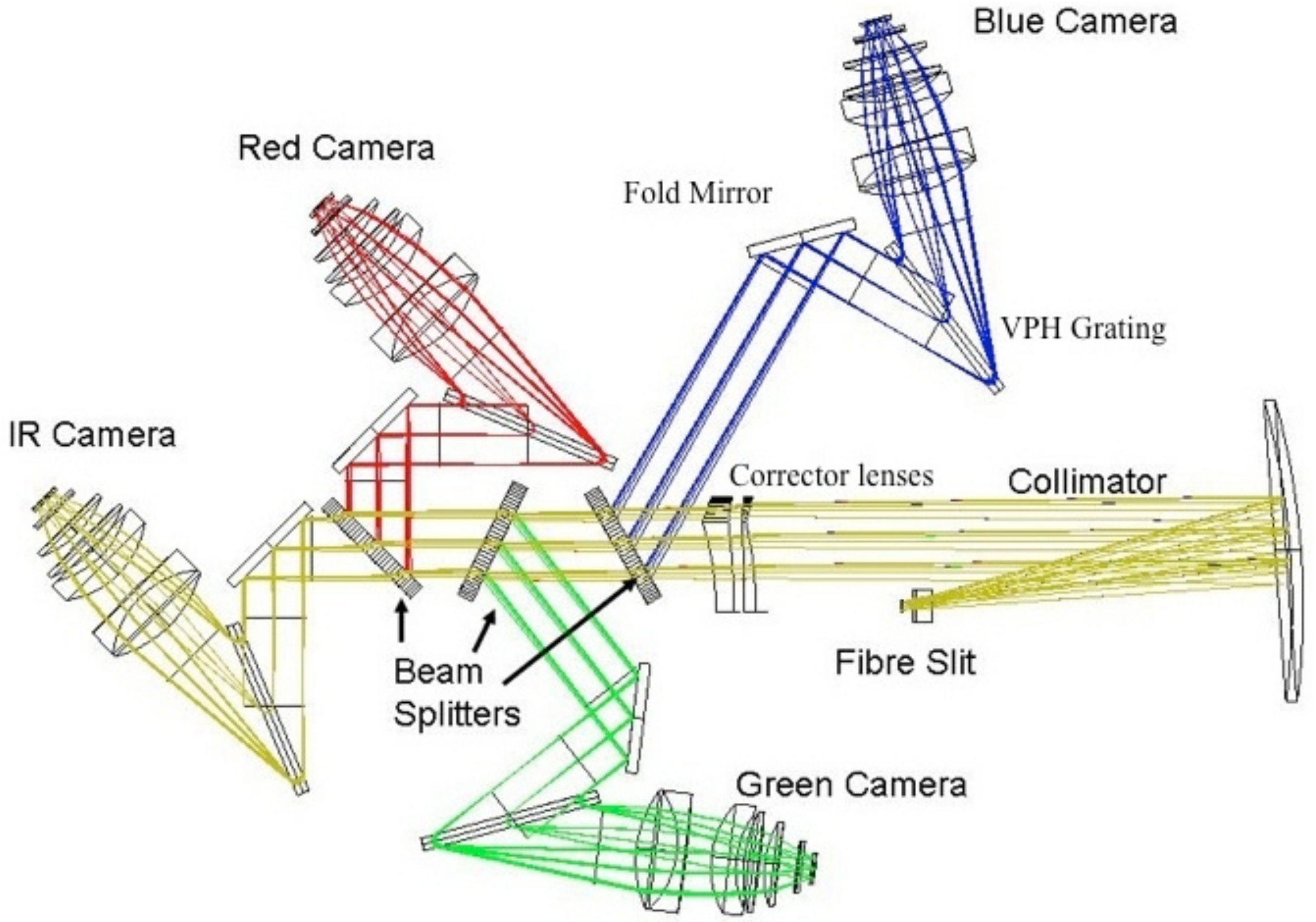}
    \caption{The optical layout of the HERMES spectrograph. Light enters at the fibre slit and then is reflected back to the collimator. Beam splitters divert light for the blue, green and red cameras, with the remaining light going to the infrared camera.}
    \label{fig:light_path}
\end{figure}

There are three main components to the 2dF/HERMES system: the 2dF fibre positioning robot and its related optics; the fibre cable that takes light from the focal plate to HERMES; and HERMES itself.

For a detailed overview of the 2dF robot and its related optics see \cite{Lewis2002}. In its current iteration it is able to position up to 392 science fibres and eight fiducial fibres (these provide guiding information for the telescope control system) onto a 2\degr-diameter circular field. The fibres are positioned sequentially around the circumference of the plate.  Of the 392 science fibres, it is usually recommended that 25 of these fibres are used for determining the sky background brightness. In the HERMES fibre cable, $\sim6$ fibres of the 392 are unavailable for use. This means that for any given configuration there are about 360 fibres available for stellar targets.

2dF uses a prime-focus corrector system, which is designed to correct for the atmospheric dispersion that would otherwise occur across the $2\degr$ field of view. This consists of a four-element corrector with two counter-rotating prismatic doublets and two other fixed lenses, mounted a few metres below the field plate when the telescope is in the observing position. 

Because the 2dF robot positions fibres one at a time, it can (depending on the complexity of the field) take anywhere from 10--50~minutes to reposition the 400 fibres for a new field configuration. In order to allow efficient use of the telescope time there are two plates. While one field configuration is being used for observations, the other plate can be configured for subsequent observing. The whole 2dF top-end is then ``tumbled'' to swap the plates from the observation position to the configuration position, and vice versa. The two plates are known as Plate 0 and Plate 1, and they each have their own sets of 400 fibres, giving a total of 800 fibres for the 2dF/HERMES system. 

The fibres are attached to the metallic field plate via rare-earth magnets housed inside a steel button which has a vertical fin for the positioning robot to grasp. The fibres themselves run parallel to the plate, necessitating a prism (with total internal reflection angle of approximately 90\degr) on each button to redirect the light that has arrived perpendicular to the plate onto the fibre core.

To provide a stable environment for HERMES, it is located several floors below the telescope prime focus in the West Coud\'{e} laboratory of the AAT building. This requires almost 50 metres of fibre-optic cable to take the light from the 2dF to the spectrograph (for AAOmega the fibre cable is 38 metres), all of which must be handled and stored when 2dF is not in use. In order to minimize focal ratio degradation the cables have a large bending radius, protective sleeves and low stress points where they are anchored. Measurements during their commissioning found that they had an average end-to-end throughput of 62 per cent at 505~nm.

The individual fibres are in groups of ten, each of which is known as a retractor at the 2dF end of the system. The full retractor assembly contains a pair of pulleys to maintain a small tension on the fibres so they remain parallel to the plate and avoid them bending when it is being moved by the robot.

Each group of ten fibres terminates inside HERMES in a block of 10 small V-grooves known as a slitlet. Each slitlet has a set of three magnifying lenses to change the focal ratio of the fibres before their light reaches the collimator.

These optics flip the order of the fibres for each slitlet, which results in a different fibre order for the plate compared to the fibre order on the slit of the spectrograph. In the nomenclature of the AAT the position on the plate is referred to as the pivot number, while the position on the slit and image is referred to as the fibre number. See Table \ref{table:fibretopivot} for the mapping between fibre number and pivot number. Unless otherwise explicitly stated, in this paper fibre numbers are those as observed by the slit.

These slitlets are grouped together into a slit block, with one slit block for each plate. These slit blocks are moved vertically inside HERMES when 2dF is tumbled. The slit block of the new observing plate is moved into the observing position (i.e., its light enters the spectrograph). The slit block of the new configuration plate is moved into a position where its fibres can be back-illuminated by LEDs to aid the 2dF fibre positioner in locating the fibres. See \cite{Brzeski2011} for more details on the design of the slitlet and fibre cable.

For details on the design and first light results of HERMES see \cite{Barden2010,Heijmans2011,Brzeski2012,Zheng2013,Sheinis2014} and references therein. HERMES produces spectra with a nominal spectral resolution of 28,000 for four non-contiguous wavelength bands, known as the blue, green, red and infrared. For the wavelength coverage, see Figure \ref{fig:bandpass_comparison} and Table \ref{table:wavelength_comparison}. The optical light path of HERMES is shown in Figure \ref{fig:light_path}. The wavelength bands were chosen to provide coverage of spectral features that could be used for determining chemical abundances that are crucial for the chemical tagging aims of GALAH. The light is split by three large beam splitters and dispersed via four large volume phase holographic gratings onto the 4096x4096 CCDs in four independently operated cameras.

\section{Data reduction}\label{sec:datareduction}

The raw spectra used in this analysis were acquired over 137 nights on the 3.9-m AAT using HERMES  as part of the ongoing main GALAH survey (January 2014 to September 2015) and one night from the GALAH pilot survey (16 November 2013). As described in Section \ref{sec:intro}, these observations are of stellar targets typically with V magnitudes between 12 and 14. The GALAH observing strategy is to observe fields when they are $\pm1$ hour of transiting the meridian, with the aim of minimizing the change in airmass over the exposure time. The fibre assignment is found using the AAO's \textsc{configure} software \citep{Miszalski2006}, which uses a simulated annealing algorithm to find the optimal configuration. There is no preference as to whether a non-fiducial fibre will be assigned to a science target or a sky position.

Usually three 20-minute exposures are acquired for each field with the aim of a signal-to-noise of 100 per resolution element in the red camera for the faintest objects. For each 2dF field configuration GALAH also acquires observations of four ThXe arc lamp for wavelength calibration, and a flap flat from a quartz 75~W lamp for determination of the position of spectra on the raw image (referred to here as a flap flat to distinguish it from other flat types that are discussed in Section \ref{sec:throughputs}).

The reduction of the raw images for this paper were performed using \textsc{matariki}, written by the lead author, which is a Python wrapper around the AAO's \textsc{2dfdr} reduction software \citep{2015ascl.soft05015A}. \textsc{2dfdr} is an automatic data reduction pipeline dedicated to reducing multi-fibre spectroscopic data, specifically the multi-fibre instruments of the AAO: AAOmega (fed by 2dF, KOALA-IFU, SAMI Multi-IFU or older SPIRAL front-ends), HERMES, the original 2dF spectrograph, 6dF, and in the future the TAIPAN spectrograph. It has evolved significantly over the years, and we have used version 6.14 for these reductions.

 \textsc{matariki} provides the interface between \textsc{2dfdr} and the raw data, instructing \textsc{2dfdr} on which exposures need to be reduced together. It makes use of the newly implemented \textsc{aaorun} command line functionality of \textsc{2dfdr} which allows for up to 28 simultaneous instances of the reduction process, greatly speeding up the data reduction for large surveys.
 
The raw images were bias subtracted using their overscan regions. Daily bias frames are acquired but investigations have shown that there is very little structure to the biases of the HERMES cameras, and that the overscan is sufficient for bias subtraction. Each extracted spectrum was divided by its equivalent flap flat extracted spectrum. Sky subtraction of the stellar spectra used the median spectrum of the 25 sky fibres in each configuration, and with the assumption that the sky brightness was uniform across the $2\degr$ field of view. Within the \textsc{matariki} wrapper, the individual spectra of a given star were combined with a simple sum. A number of targets were observed on more than one night but only those spectra observed as part of a continuous observing block were combined together.

There were a total of 118,881 co-added spectra in our final dataset. We have excluded spectra with poor wavelength solutions, or those stars for which there was no AAVSO Photometric All-Sky Survey (APASS) photometry. Due to a general observer bias to start a night's observations on Plate 0, there are slightly more Plate 0 spectra (51.3 per cent) than Plate 1 spectra. Some fibres have been disabled for the entire survey. On Plate 0 there were 6 disabled fibres (Fibres 66, 141, 188, 307, 357, 377) and on Plate 1 there were 7 disabled fibres (Fibres 1, 24, 47, 60, 133, 190, 262).

Since we are using spectra that were taken in a variety of seeing and atmospheric conditions, it is important that there be a sufficient number of spectra per fibre for our results to be statistically significant. Excluding the disabled fibres, for Plate 0 the median number of spectra per fibre was 157 and for Plate 1 it was 147. Even with 400 possible stars per field, some GALAH field configurations do not use all of the fibres, and some fibres were disabled for short periods of time. For Plate 0 there were 33 fibres with fewer than 150 spectra, of which only one fibre had fewer than 100 spectra (Fibre 110 with 93 spectra). For Plate 1 there were 31 fibres with fewer than 140 spectra, of which two fibres had fewer than 90 spectra (24 on Fibre 4 and 83 on Fibre 139). Overall, of the 771 fibres with usable data, all but one had at least 83 spectra.

\section{Throughput determination}\label{sec:throughputs}

There are external factors independent from the fibres and spectrograph themselves that will impact the flux that is received by the detector. One factor is the fibre positioning accuracy. For a detailed discussion of the effects of fibre positioning errors in fibre-fed spectrographs see \cite{Newman2002}. The fibres must be positioned accurately on the plate in order for the desired target's light to be fall onto the fibre core. For this it is necessary to determine the coordinate transformations between the sky position and the plate position. This is performed at the AAT via a procedure known as POSCHECK where stars with well known positions are observed with a focal plane imager. This is done every time 2dF is remounted on the top end of the telescope. Although the POSCHECK is done as accurately as possible by the astronomers and technicians at the AAT, there will still be small errors in positioning the fibres. 

Other external factors that will affect the perceived transmission of the fibres relate to the Earth's atmosphere. The seeing will cause the stellar image to change size, affecting the amount of light that is received by the fibre. There are also variations in the overall atmospheric transmission. It is assumed that with the large dataset these effects will be averaged out for a given fibre.

To determine the relative throughputs of the fibres we compare the expected brightness of each star with its recorded flux on the detectors. Unfortunately there are no all-sky photometric surveys with filter sets that perfectly capture the wavelength ranges of the gratings of HERMES. As such it is necessary to use close equivalents. We make use of the GALAH input catalogue's photometry acquired by the APASS. For the blue, green, red, and infrared cameras we are using their measurements from Johnson B \& V filters and Sloan r \& i filters, respectively. The overlaps between the wavelength coverage of the photometric filters and that of the cameras and gratings of HERMES are given in Figure \ref{fig:bandpass_comparison} and Table \ref{table:wavelength_comparison}.

For each reduced, co-added spectrum the mean flux count per pixel per second of exposure was determined: $F_\textrm{X}$, where X refers to the blue, green, red, and infrared cameras, which are shortened to B, V, R, IR respectively in this work. This was then converted to a magnitude via the elementary equation,
\begin{equation}
	\textrm{mag}_{\textrm{spec,X}} = -2.5\log_{10}(F_\textrm{X}),
\end{equation}
The difference from the APASS photometry was then found,
\begin{equation}
	\Delta \textrm{mag}_{\textrm{X}} = \textrm{mag}_{\textrm{APASS,X}} - \textrm{mag}_{\textrm{spec,X}}.
\end{equation}
Finally this was converted to a flux ratio and normalized to the median of all the flux ratios for that camera
\begin{equation}\label{eq:deltamagX}
	\deltamagX = \frac{10^{\Delta \textrm{mag}_{\textrm{X}}}}{\textrm{median of all}(10^{\Delta \textrm{mag}_{\textrm{X}}})}.
\end{equation}
So the relative fibre throughputs for the blue, green, red, and infrared cameras are \deltamagB, \deltamagV, \deltamagR, and \deltamagIR\ respectively.

Apart from the blue camera, the wavelength coverage of the HERMES's cameras lies mostly within the equivalent photometric filter. Since we are interested in the relative behaviour of the fibres, the fact that our photometric bandpasses do not correspond perfectly to the grating wavelengths is not an impediment. We are not trying to determine the absolute throughput of the system. The key is using photometry that samples the star light at approximately the correct wavelengths in order to provide a good estimate of the brightness of the stars in the HERMES bandpasses. Due to the range of stellar temperatures observed by GALAH, simply using (for instance) a scaled V magnitude would cause incorrect estimates of the spectral brightness in the IR camera of cool and hot stars.

\subsection{Twilight flat analysis}
As discussed at the start of this section, stellar observations can have problems related to fibre positioning and atmospheric variation. Observations of a uniform flat light source avoids these disadvantages. Conversely, while flat field observations are useful for understanding intrinsic fibre-to-fibre variations, they would mask throughput variations due to systematic positioning errors. Although positioning errors are not actually the fault of the fibres themselves, they do introduce variations in the signal recorded by each fibre for stellar targets, which is the `real world' for these fibres.

There are three different flat calibration observations that are acquired by GALAH: flap flats, twilight flats, and dome flats. The flap flats use a white light lamp that illuminates large flaps that are mechanically inserted into the telescope beam between the focal plane and the mirror. The flap flat exposures (90 -- 180 seconds for GALAH) are acquired for each fibre configuration and their primary use is for determining the location of the fibre traces on the detectors. Although bright, the illumination of the flaps is uneven and the light path very dissimilar to that of astronomical targets as the light does not come from the telescope mirror. This means that individually they are not useful for fibre-to-fibre throughput mapping.

Twilight flats are observations of the bright twilight sky, usually with exposures of about 120 seconds. This illuminates the entire focal plane, and each fibre produces a solar spectrum superimposed with a sky spectrum. The procedure for GALAH is to point the telescope $15\degr$ east of the zenith. \cite{Chromey1996} shows that this is (broadly) the best position to point a telescope in order to minimize gradients across the field of view, especially important for the 2\degr\ field of 2dF. A small number of dome flats have also been acquired. Dome flats take much longer to acquire adequate signal with HERMES ($\sim 30$ minutes) so there have not been enough taken with a variety of 2dF configurations to make them worthwhile for this fibre throughput analysis.

A twilight flat has the advantage over a dome flat that its light path is more similar to that of an astronomical target, as the light is coming from "infinity", rather than from a screen inside the dome.

For the analysis of twilight flats, it is not possible to easily determine what absolute sky flux was incident on the telescope. Unlike the stellar observations for which we have photometry, with twilight flats we do not have any straighforward method for estimating how many photons the fibre was meant to have received. The sky brightness varies due to differing atmospheric clarity, where in the sky the telescope is pointing, and time since sunset or before sunrise. Some of the twilight flats in the GALAH dataset were taken up to 25 minutes after sunset, meaning the sky brightness was very low. Only those twilight flats that were acquired within 12 minutes of sunset were used in this work (there were no useful morning twilight flat exposures). We have made the assumption that there is no gradient across the focal plane and all fibres received the same uniform amount of light.
 
For Plate 0 there are typically $\sim40$ twilight flat spectra per fibre and for Plate 1 there are fewer: $\sim24$. Again, this discrepancy in the number of spectra per plate is because of the preference to start with Plate 0 at the start of the night. Twilight flats will typically use the first GALAH field configuration of the night, so the fibres are randomly distributed across the field.

The twilight flats were analysed in the following way. They were reduced using \textsc{2dfdr} as if they were stellar observations (i.e., reduced with flap flat and arc exposures to determine the tramline maps and wavelength solution), but with no sky subtraction (since the fibres allocated as sky fibres should contain exactly the same signal as the `star' fibres in a twilight flat, sky subtraction would result in `zero' flux in all fibres).

As with the stellar spectra, the mean counts of each twilight spectrum was found. Because of the variation in sky brightness it was necessary to normalize each twilight exposure (e.g., the 360 spectra on each detector that were acquired at the same time) by the mean fibre count for each exposure. This normalized mean count is the \deltamagX\ for the twilight exposures.

\section{Results}\label{sec:results}

\begin{figure*}
	\includegraphics[width=\textwidth]{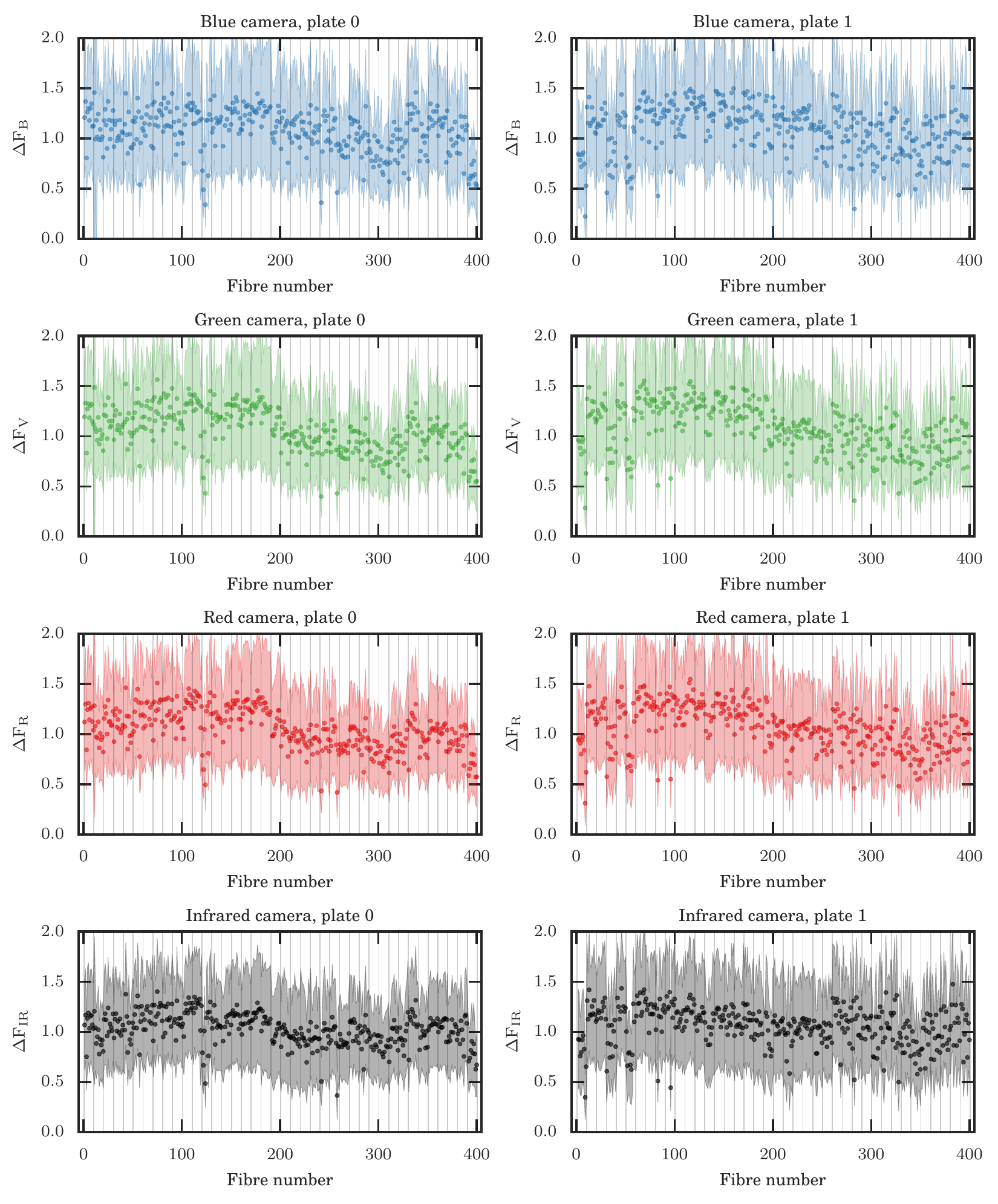}
    \caption{The mean relative throughputs of each fibre for each of the four cameras on each of the two plates. The dots give the mean and the shaded areas are one standard deviation about this mean. \deltamagB, \deltamagV, \deltamagR, and \deltamagIR\ were determined by Equation \ref{eq:deltamagX}. The black vertical lines on each plots divide fibres into their respective retractor groups.}
    \label{fig:mean_throughputs}
\end{figure*}

\begin{figure}
	\includegraphics[width=\columnwidth]{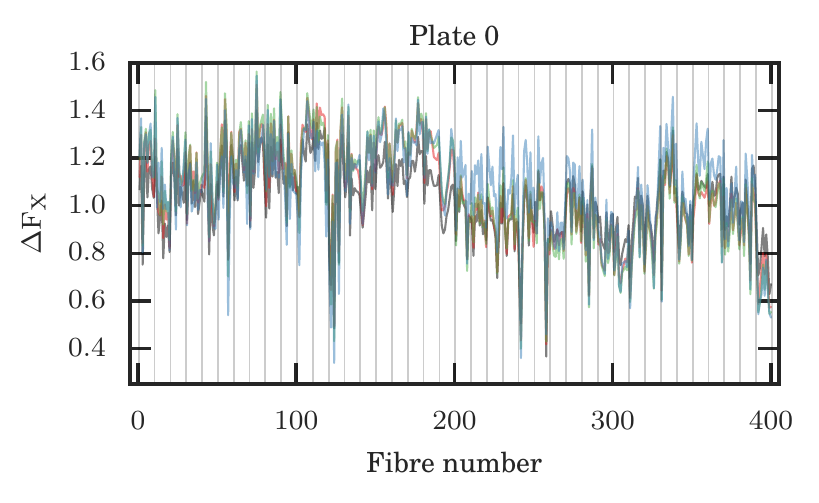}
	\includegraphics[width=\columnwidth]{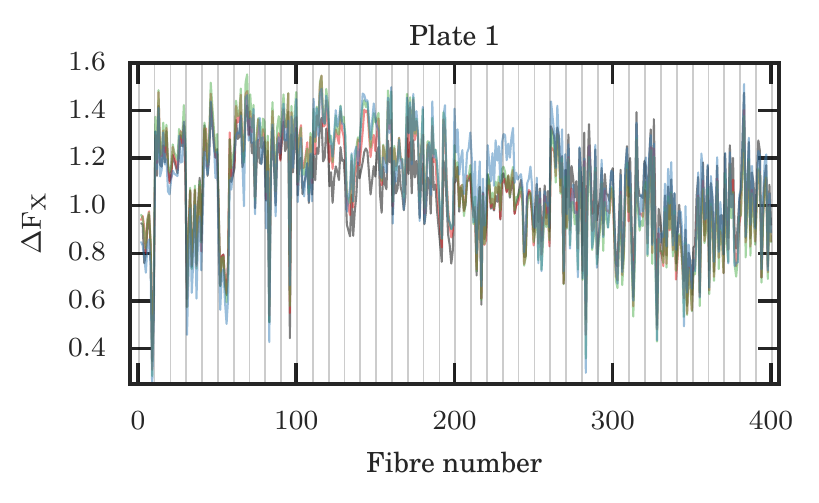}
    \caption{The same mean relative fibre throughputs as Figure \ref{fig:mean_throughputs} but for each plate, the results from each camera are overplotted. Here \deltamagX\ refers generically to the throughputs of each camera (\deltamagB, \deltamagV, \deltamagR, \deltamagIR). This confirms that there is some camera-to-camera variation, but the overall fibre-to-fibre profiles are similar for each camera.}
    \label{fig:mean_throughputs_all}
\end{figure}

\begin{figure}
	\includegraphics[width=\columnwidth]{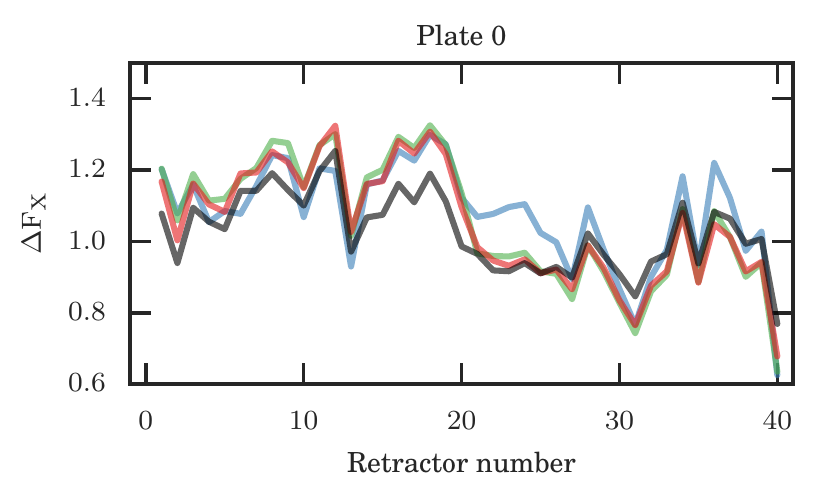}
	\includegraphics[width=\columnwidth]{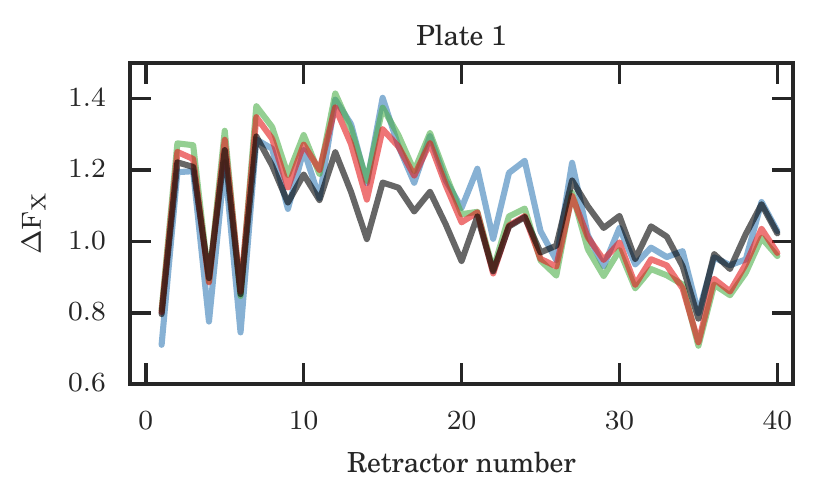}
    \caption{The mean relative throughputs of each retractor for each of the four cameras on each of the two plates. Here \deltamagX\ refers generically to the throughputs of each camera (\deltamagB, \deltamagV, \deltamagR, \deltamagIR) as determined by Equation \ref{eq:deltamagX}. The results from the four cameras are plotted together. This indicates that for both plates there is a general decline in the relative throughputs with increasing retractor number. There is also some chromatic variation especially for the middle retractors. On both plates from about retractor 13--20, the infrared camera has systematically lower throughput, while for retractor 20--27 it is the blue camera that is systematically higher than the other cameras.}
    \label{fig:retractors}
\end{figure}

\begin{figure}
	\includegraphics[width=\columnwidth]{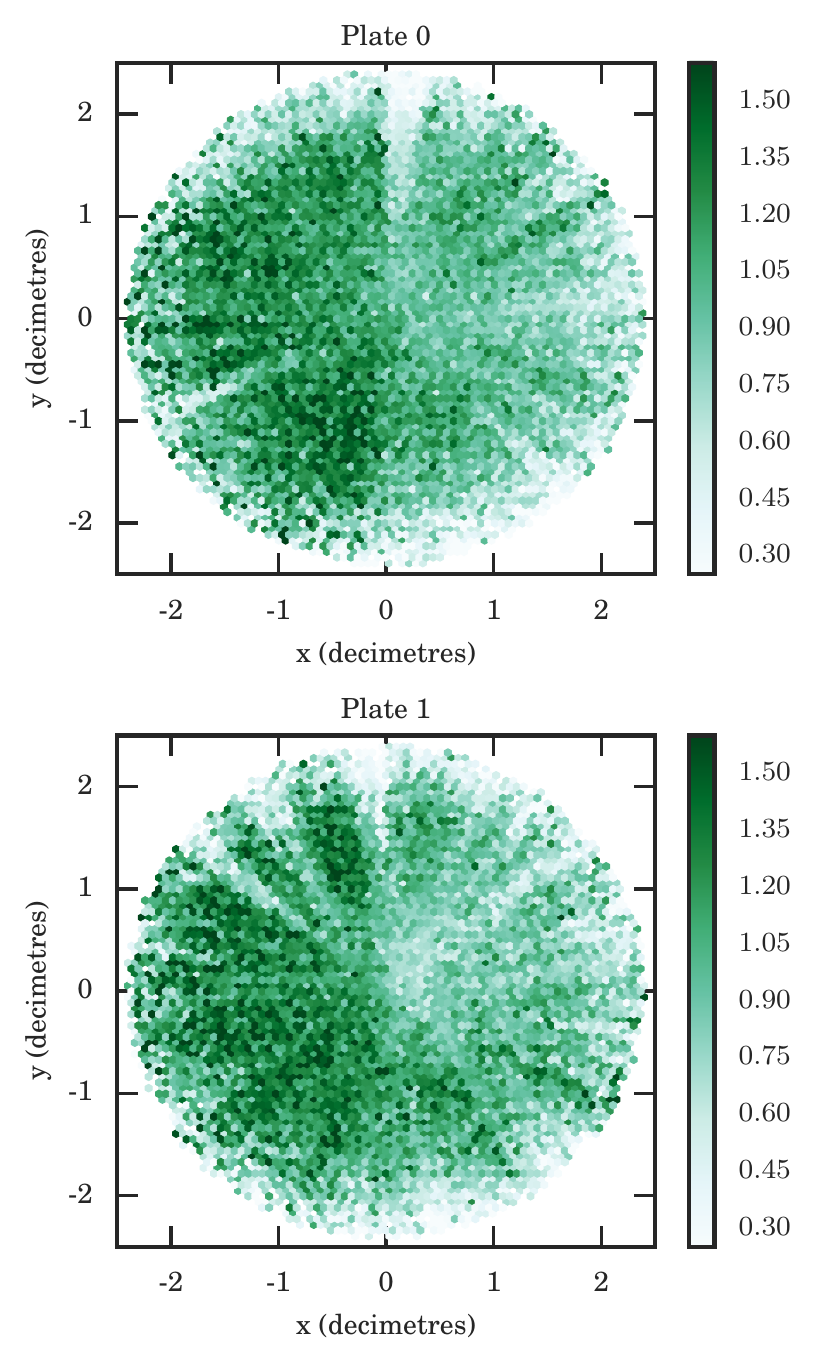}
    \caption{The relative fibre throughputs for different physical fibre positions on each plate. North is to the right and east is towards the top. Each small hexagonal section of the plate region is coloured to indicate its mean relative throughput in the green camera (\deltamagV) for that section of the plate. The x and y positions are taken from the FITS header of the raw image. Typically (for GALAH fields) each fibre would be found in a segment of the plate area. This is why there are low throughput segments on the plate: these indicate regions where the fibres of low-throughput retractors are typically placed. The pivot number (the fibre order on the plate) increases counter-clockwise, with pivot 1 at the 12 o'clock position on each plate (the map of pivot to fibre number can be found in Table \ref{table:fibretopivot}). For radial distances larger than 2 decimetres, the number density of spectra in each hexagon falls below five, while at the centre of the plate they contain about 25 spectra.}
    \label{fig:plate_hex}
\end{figure}

Figures \ref{fig:mean_throughputs}, \ref{fig:mean_throughputs_all}, \ref{fig:retractors}, and \ref{fig:plate_hex} present the throughputs found from the stellar observations. In Figures \ref{fig:mean_throughputs} and \ref{fig:mean_throughputs_all} these results are shown averaged per fibre per plate, in Figure \ref{fig:retractors} they are averaged per retractor per plate, and in Figure \ref{fig:plate_hex} the results are binned and averaged for spatial positions on each plate. There are a number of key findings: on both plates the first 200 fibres appear to have a higher throughput than the last 200 fibres; there are some small chromatic variations in the throughputs for some fibres; each plate has a number of retractors/slitlets that show poor throughput; and a small number of fibres have much lower throughputs than the rest of the fibres.

\subsection{Large-scale variations}\label{sec:leftright}

\begin{table}
\caption{The gain and bias levels of the amplifiers of the four HERMES cameras. The gain values are taken from the header of the image and the bias levels were taken from a representative bias image.}
  \label{table:bias_gain}
  \begin{tabular}{lcccc}
  \hline
  & \multicolumn{2}{c}{Gain (e$^{-}$/ADU)} & \multicolumn{2}{c}{Bias (counts per pixel)} \\
   Camera & Amplifier 1 & Amplifier 2& Amplifier 1 & Amplifier 2\\
    \hline
Blue & 2.70 & 2.60 & 283 & 292\\
Green & 3.01 & 2.99 & 249 & 305\\
Red & 2.83 & 3.17 & 168& 226\\
Infrared & 2.75 & 2.50 & 314 & 426\\
   \hline
  \end{tabular} 
\end{table}

\begin{table}
\caption{The mean relative throughputs (as determined by Equation \ref{eq:deltamagX}) found for each half of the plate. They are remarkably stable on the two halves of both plates, which indicates that the throughput variations are a symptom of fibre positioning and the robot gantries, rather than the fibres themselves.}
  \label{table:plate_half_throughputs}
  \begin{tabular}{lcccc}
  \hline
 \multicolumn{5}{c}{Stellar observations}\\
  & \multicolumn{2}{c}{Plate 0} & \multicolumn{2}{c}{Plate 1} \\
   Camera & First 200 & Last 200  & First 200 & Last 200 \\
    \hline
Blue & $1.16\pm0.62$ & $1.00\pm0.52$ & $1.16\pm0.63$ & $1.02\pm0.55$\\
Green & $1.20\pm0.58$ & $0.92\pm0.45$ & $1.21\pm0.60$ & $0.95\pm0.49$\\
Red & $1.18\pm0.55$ & $0.92\pm0.45$ & $1.18\pm0.58$ & $0.96\pm0.49$\\
Infrared & $1.10\pm0.49$ & $0.96\pm0.45$ & $1.11\pm0.54$ & $1.01\pm0.50$\\
   \hline  
 \multicolumn{5}{c}{Twilight flats}\\
Blue & $1.02\pm0.17$ & $0.98\pm0.18$ & $1.04\pm0.21$ & $0.96\pm0.15$\\
Green & $1.05\pm0.18$ & $0.94\pm0.17$ & $1.07\pm0.22$ & $0.93\pm0.16$\\
Red & $1.04\pm0.16$ & $0.96\pm0.16$ & $1.06\pm0.22$ & $0.94\pm0.16$\\
Infrared & $0.99\pm0.16$ & $1.01\pm0.18$ & $1.01\pm0.21$ & $0.99\pm0.17$\\
   \hline  
  \end{tabular} 
\end{table}

\begin{figure}
	\includegraphics[width=\columnwidth]{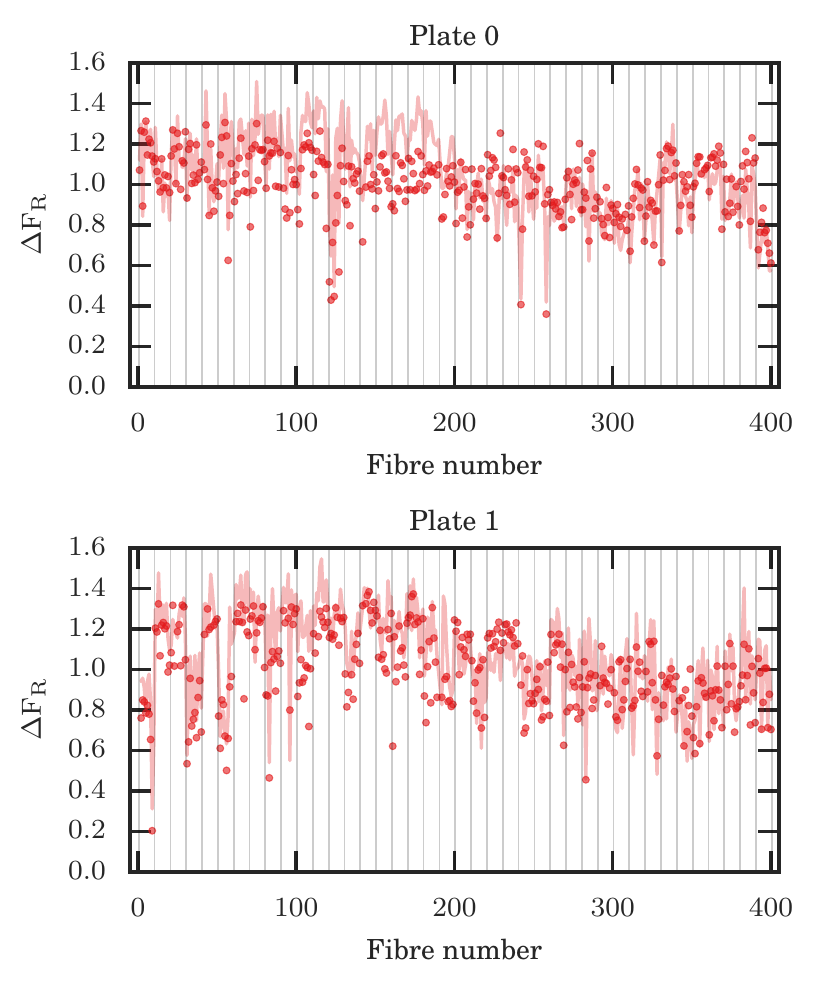}
    \caption{The mean relative throughputs for each fibre for the red camera (\deltamagR; as determined by Equation \ref{eq:deltamagX}) for Plate 0 (top) and Plate 1 (bottom) for the twilight flat exposures (red dots). Under-plotted is a line showing the equivalent values for the stellar observations. This shows that the twilight flats show much of the same behaviour such as poor retractors, bad fibres, but do not have the large drop in throughputs found on Plate 0.}
    \label{fig:twilight_red}
\end{figure}

Evident in Figures \ref{fig:mean_throughputs}, \ref{fig:mean_throughputs_all}, \ref{fig:plate_hex}, and \ref{fig:retractors} is that the fibres on the first half of each plate are giving consistently higher throughputs than those on the second half of each plate. It is more obvious on Plate 0 and due to chromatic variations (see Section \ref{sec:chromatic}), it is not as obvious for the results from the blue camera.

GALAH uses two amplifiers to read out its exposures (the HERMES detectors are capable of using 1, 2 or 4 amplifiers), with the fibre traces parallel to the serial registers. As each amplifier has a different bias level and gain (Table \ref{table:bias_gain}), if these were not being correctly handled in the reduction process, this could cause the throughput differences. However the amplifier that reads out of the first half of fibres has the lower bias level on all four cameras, so this would result in lower throughputs for the first half of the fibres. It cannot be caused by the gain as the gain of the red detector for the first amplifier is lower than the gain for the second amplifier, while it is the other way around for all the other cameras.

A possible explanation is that this is the result of fibre positioning errors. In observations with 2dF, it is crucial that the fibre is positioned accurately and precisely such that the starlight received by the detector is maximized over the (nominally) one hour total exposure time. The fibre cores are only 140~\micron\ in diameter (equating to about 2\arcsec\ on the sky), requiring micron level precision. We find that the average throughputs for the fibres from 1--200 and from 201--400 for each camera, are remarkably similar for both plates (Table \ref{table:plate_half_throughputs}). 

The two plates share the focal plane imager (FPI) gantry and fibre gripper gantry system, as these are fixed and the plates tumble between them. The FPI camera is used on the observation plate, and the gripper gantry works on the configuration plate. Whenever 2dF is placed at the top-end of the telescope, routines are performed on-sky to update the 2dF control software as to the relative behaviours of these two gantries. It is known that there is an amount of non-linear behaviour that is not taken into account with these checks but are taken into account with the \textsc{configure} software.

We can use the twilight flats to determine if the fibre placement is the cause of the variation (Table \ref{table:plate_half_throughputs} and Figure \ref{fig:twilight_red}). During a twilight (or dome) flat, the entire focal plate is illuminated, so fibre placement errors should not cause problems. The retractor-to-retractor variability is recreated and the same bad fibres can be identified. However the stellar targets have larger throughputs for the fibres from about 120 to 200 and there does not appear to be the same sharp drop in throughput for the twilights. The overall averages from the two halves of the plate show that, although there is a small difference between the first and second halves for the twilights, it is nowhere near as large as that observed for the stellar observations. 

It is worth noting that a typical co-added stellar spectrum from GALAH has mean (arbitrary) flux counts per pixel of $\sim200$ from one hour of exposures, while a twilight flat has counts in the tens of the thousands from $<500$ seconds exposure. There is some crosstalk of the fibre traces on the HERMES's detectors, which would have a larger effect for brighter spectra.

A further possibility relates to the fine guidance of the telescope. The automatic guiding of the telescope using eight fiducial fibres. As part of the field configuration process it is necessary to supply possible guide stars to which this fibres can be allocated. The \textsc{configure} fibre allocation software requires at least three fiducial fibres be allocated, and in the case of GALAH observing all the eight fibres are almost always used because of the ability to have 20-30 potential guide stars per field. These stars are chosen from the GALAH input catalogue, so they are on the same astrometric system as the science targets (actually a fundamental requirement of the 2dF guiding system).

Typically the stars are guided for a wavelength of 5000\AA\ (equating to blackbody spectrum for a 5500 K object, typical for Galactic disk stars). The guiding is automatically updated once every $\sim1$ second. The guide fibres themselves consist of 8 individual fibres, and the guiding software is aiming that the guide stars' light is primarily in the centre of these individual fibres. Since it is the full ensemble of guide stars that are being used, and these are spread across the field, we consider it unlikely that guiding problems over the course of the exposures would manifest as the results we have found.

Overall, it does seem that there are systematic positioning errors which are causing fibres on one half of each plate to be not as well aligned to the incoming light. This result was not identified by \cite{Sharp2013} because the main analysis was performed using the OH sky lines, which will not suffer from throughput variations due to positioning problems.

\subsection{Chromatic variation}\label{sec:chromatic}
In Figures \ref{fig:mean_throughputs_all} and \ref{fig:retractors} the results from the four cameras are over-plotted. There is consistent behaviour between the four cameras, e.g., the large-scale variations (Section \ref{sec:leftright}), low-throughput retractors (Section \ref{sec:retractors}) and poor individual fibres (Section \ref{sec:poor_fibres}). But there are a few differences between the four cameras, mostly clearly seen in Figure \ref{fig:retractors}. From about retractor 12 to 20 on both plates, the infrared camera is systematically lower in throughput, while the other three cameras are consistent with each other. Then from retractors 20 to 27, the blue camera is higher while the other three cameras are consistently lower. The green and red cameras never deviate from each other.

It is unclear what would be causing this chromatic variation for only specific groups of fibres. The fibres are constructed from Polymicro FBP \citep{Brzeski2011}, a material chosen by the AAO due its proven past performance. Since the chromatic effects are happening on both plates for the same retractors or areas on the plate, this would indicate that the chromatic variation is not the fault of the fibres themselves (as each plate has its own independent set of fibres).

These chromatic effects seem to be independent of the plate or physical fibre in use, so the cause needs to be something common to both plates. This could be the optical elements before or after the fibres themselves in the light path.

In Section \ref{sec:design} we discussed the prime-focus corrector system. It is possible that there is some previously unidentified chromatic variation in this system. However why it would affect only one area of the plate is unclear. The elements of the prismatic doublet are in different positions for different telescope pointings. Due to the large number of possible target fields, GALAH observations are typically made when the objects are $\pm1$ hour of their transit time. This could result in the pupils of the fibres always seeing the same portions of the prime-focus corrector system.

Inside HERMES it is possible that there could be some previously unidentified chromatic effects related to the gratings or field flatteners. During the construction of HERMES it was found that the VPH gratings for the green and red cameras had poor wavefront performance, which required post-polishing. It is curious that these are the two cameras which in fact do not deviate.

It is possible that these chromatic effects are the results of a combination of factors, and that the blue and the infrared deviations are caused by different parts of the light path.

\subsection{Retractors with poor throughputs}\label{sec:retractors}
Figures \ref{fig:retractors} and \ref{fig:plate_hex} show that some of the retractors on each plate have much lower throughputs overall compared to the rest of the retractors. The results from \cite{Sharp2013} similarly show strong retractor-to-retractor variations on each plate (see their figure 7 and section 7.6), so this is not unique to HERMES. Unfortunately it is difficult to disentangle the two ends of the HERMES fibre cable to determine whether this is the result of some problem with the retractors at the 2dF end, or with the slitlets at the HERMES end.

For Plate 1, retractors 1, 4, 6 and 35 are especially low efficiency, and their traces can be observed in Figure \ref{fig:plate_hex} as white radial regions. They are about a third lower in throughput than the surrounding retractors (which would equate to making the stars half a magnitude fainter). For Plate 0, there are fewer poor retractors: retractors 31 and 40 has consistently poor fibres, but it is unclear if retractor 13 is a poor retractor or just an unlucky to have a group of poor fibres. In Figure \ref{fig:mean_throughputs} it can be observed that only some of its fibres have low throughput, unlike for the previously identified retractors in which all the fibres exhibit poor throughputs. 

As the fibres of the retractors are typically fabricated together, it is possible that a poor retractor is the result of environmental factors during the manufacturing process (e.g., changes in temperature, pressure, humidity). Another possibility is that it is not the fibres themselves that are at fault, but in fact the slitlet inside HERMES, such as a misalignment of its mounting on the slit block or its magnification optics. This could manifest itself as poor resolving power for these fibres and retractors. 

ThXe arc lamp exposures are acquired with each set of science exposures. The reduction software produces a one-dimensional arc spectrum for each fibre with the same extraction and reduction methods as the science exposures. Gaussians were fitted to more than 20 arc lines in each fibre in each arc exposure to determine their width (in the form of $\sigma$ of the Gaussian), line position and amplitude. The PSF of HERMES varies across the camera, such that the arc lines well away from the centre of the detector are not perfectly Gaussian in shape, however for simplicity Gaussians were used for all of the arc lines.

\begin{figure*}
	\includegraphics[width=\textwidth]{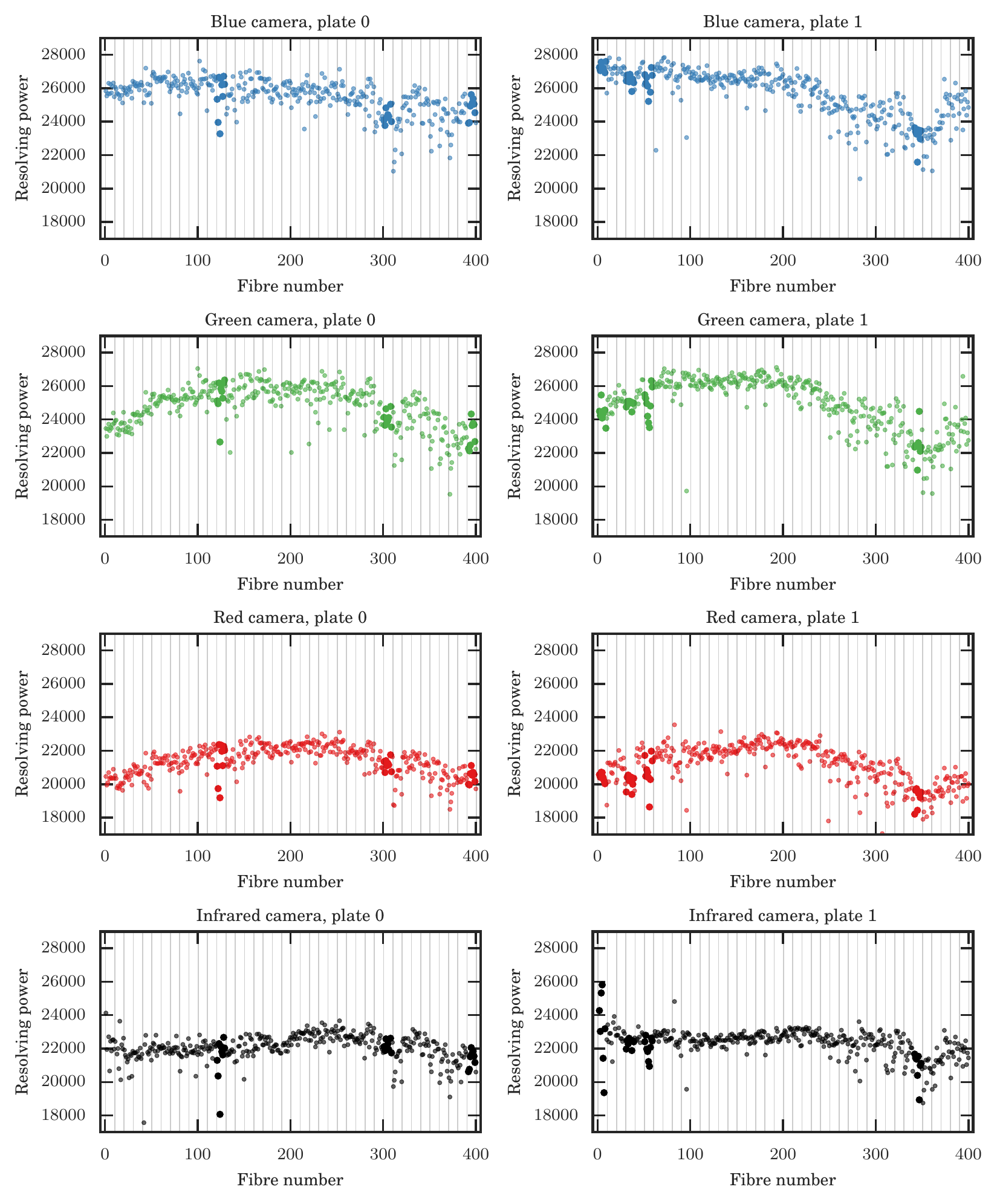}
    \caption{The average resolving powers of each fibre for each plate as determined from Gaussians fitted to multiple arc lines in each reduced arc spectrum. This resolving power is derived from the width of the arc lines in the ThXe spectrum, and is therefore the convolution of the arc line widths and the instrumental profile. Large dots are used for fibres from the retractors that were identified as having overall low throughputs. In some cases these retractors appear to coincide with lower resolving power, but not in all cases.}
    \label{fig:resolutions}
\end{figure*}

The $\sigma$ of each arc line in each fibre was converted to a resolving power ($R = \lambda / [2\sigma\sqrt{2\ln2}]$) and these resolving powers from all the arc lines were averaged for each fibre on each plate. The results are shown in Figure \ref{fig:resolutions}. The nominal spectral resolving power of HERMES is 28,000. As these results come from fitting Gaussians to likely non-Gaussian PSFs, they should not be taken as absolute values. It should be remembered that resolving power values are the result of the convolution of the instrumental profile and non-delta function arc lines, with an assumption that they can be modelled with a Gaussian. The arc lamp illumination of the fibres is also done via the flaps, so the fibres are not illuminated as they would be for an astronomical object. On each plate the resolving power profiles for the blue \& infrared cameras are similar and likewise for the red \& green cameras. This is probably caused by the optical and mechanical design of HERMES and manufacturing of the VPH gratings.

On Figure \ref{fig:resolutions} the retractors on each plate that were identified as low throughput are marked with larger dots. For Plate 1, retractors 4, 6 and 35 correlate with regions of lower resolving power. Likewise on Plate 0, retractors 13 and 31 have lower resolving power. However retractor 40 on Plate 0 and retractor 1 on Plate 1 do not. It is clear that some of the low-throughput retractors correspond to poorer resolving power but not always. In the case of retractor 13 on Plate 0, we find that some of the fibres have low resolving power, while some look unaffected, mimicking the results observed for the throughputs, where only some of the fibres have low throughputs.

The resolving power results indicate that for some of the retractors with low throughputs, the culprit appears to be the slitlets inside HERMES. The low resolving power and low throughputs would be the manifestation of slight misalignments of a slitlet or its optics.

\subsection{Individual low throughput fibres}\label{sec:poor_fibres}
As evident in Figure \ref{fig:mean_throughputs}, there are a number of fibres which have low throughputs, independent of their retractor. However, the retractor-to-retractor variation makes it hard to quantitatively determine bad fibres from just their mean \deltamagX.

Combining the results from Figures \ref{fig:mean_throughputs} and \ref{fig:retractors}, the mean \deltamagX\ for each fibre was divided by the mean \deltamagX\ of its retractor, giving \deltamagX$_{\mathrm{,norm}}$. This gives a much more stable result around unity that could be used to identify bad fibres. The standard deviation of \deltamagX$_{\mathrm{,norm}}$ was about 15 per cent for the four cameras on both plates. Fibres which were consistently \deltamagX$_{\mathrm{,norm}}>\pm|30\%|$ from unity on a given plate across the four cameras were flagged. For Plate 0 this identified seven pivots which were lower than average (54, 127, 129, 249, 253, 286, 340) and two pivots which were higher than normal (48, 122). For Plate 1 there were ten pivots (2, 40, 88, 95, 214, 262, 288, 318, 323, 356) and six pivots (53, 302, 316, 325, 327, 388) respectively. Pivot 122 on Plate 0 comes from the problematic retractor 13 which gave a large range of fibre throughputs which caused issues for normalizing by its retractor mean. This retractor also features at least two pivots that would be low throughput if its retractor value was correct.

With 386 usable fibres on Plate 0, (and including the two extra poor fibres from retractor 13), 2.3 per cent of fibres are giving throughputs that are significantly below the average behaviour of their retractor. For Plate 1, with 385 usable fibres, 2.7 per cent of fibres are poor when their retractor's overall throughput is taken into account. Across both plates, this equates to 2.5 per cent of all usable fibres. \cite{Sharp2013} found that about 5 per cent of the fibres in the old AAOmega fibre cable were `stable low', meaning they were consistently below 90 per cent for their measure of the transmission of the fibres.

It is likely that the poor HERMES fibres are the result of misaligned prisms, build up of dirt on the prism surface, and/or small air gaps between the prism and the fibre.

\subsection{Temporal degradation}
\begin{figure}
	\includegraphics[width=\columnwidth]{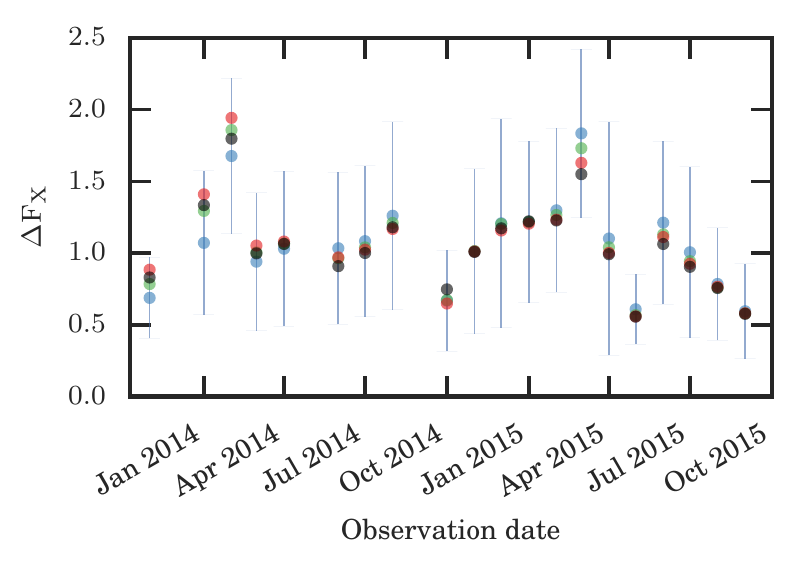}
    \caption{Mean relative fibre throughputs (\deltamagX; as determined by equation \ref{eq:deltamagX}) for each month across both plates, coloured for each camera. The error bars give the scatter for the blue camera, and are indicative of the other cameras. Although there is some month-to-month scatter, there does not appear to be any temporal degradation of the fibres. The months with the highest average throughputs (February 2014 and March 2015) actually consist of only one observing night, and so have the least amount of data.}
    \label{fig:time_series}
\end{figure}

\cite{Sharp2013} looked at the individual fibre histories to identify any long-term trends, such as fibres that were stable, slowly declining, step functions (fibres with sudden increases or decreases in throughput, likely from repairs or damage), variable histories, rising throughputs, or the aforementioned `stable low'. On inspection of individual fibre histories for the GALAH dataset, all of the HERMES's fibres exhibited stable throughput histories, with no evidence for step functions, increases or declines. There are some fibres that have been disabled after some use in the survey (e.g., pivot 101 on Plate 0 which has no observations from March 2015 onwards).

As well as individual fibre histories, it is useful to see if there has been any overall degradation in the fibre throughput over all of the fibres. Although the timespan of the GALAH survey has been short compared to the dataset used in \cite{Sharp2013} ($\sim7$~years), it was prudent to check if there had been any degradation in fibre quality over the course of the survey. The results of Figure \ref{fig:time_series} would indicate that there has not been. All of the results, regardless of plate, have been grouped by calendar month and averaged. The month-to-month scatter is the result of the combination of the factors: some months have few observing nights (November 2013 had only 345 measurements in this dataset while February 2015 had nearly 13,000) and there is variable seeing and weather quality. It is likely that consecutive nights will suffer or enjoy the same quality of seeing and weather.

\section{Concluding remarks}\label{sec:discussion}
All fibre-fed instruments exhibit some amount of fibre-to-fibre variations in throughputs. This is part of the difficulty of sky subtraction with a fibre-fed spectrograph; namely that you need to understand the throughput variations in order to correctly combine and scale your sky spectra to subtract from the science fibres \citep[for a discussion of sky subtraction with fibre-fed instruments, see][]{Wyse1992}.

There is surprisingly little in the literature of similar investigations on other multi-object, fibre-fed spectrographs apart from the aforementioned \cite{Sharp2013}. \cite{Fabricant2005} reported that for the HectoSpec instrument there was very stable throughput (2 per cent standard deviation) for the fibres from laboratory measurements. They however do not report values for stellar targets. \cite{AllingtonSmith2002} "showed an rms variation of 6 per cent" for the Gemini Multiobject Spectrograph (GMOS), again from non-astronomical targets. \cite{Kimura2010} found that fibres of Subaru's Fibre Multi-Object Spectrograph (FMOS) had a variability in throughput of about 30 per cent for a bright black body source (which they note is different to an on-sky test).

Some of the effects identified in the HERMES results may not be present in other spectrographs and positioners. Although it has been upgraded, 2dF is a relatively old instrument, and newer instruments have been built with the knowledge of aspects of 2dF that did or did not work. Also, its particular method of fibre placement (serially with an x/y robot gantry) is also not the only option available: there are `echidna' positioners \citep[e.g., as used by FMOS on Subaru;][]{Akiyama2008}, plug plates \citep[e.g., as used by the APOGEE spectrograph;][]{Wilson2010}, and `Starbugs' \cite[e.g., to be used as part of the AAO's new TAIPAN spectrograph;][]{Gilbert2012}. Overall, we believe it would be extremely useful to undertake analysis of the relative throughputs of spectrographs using their targets of interest, rather than in the laboratory with light sources that may not be representative of the real world conditions that the fibres and spectrographs will encounter.

There are some recent and upcoming changes to 2dF and HERMES that could affect the relative throughputs of the fibres. The first change was in November 2015 when one of the 2dF plates was replaced with a new plate constructed from Invar, a nickel-iron alloy noted for its very low coefficient of thermal expansion. Since its inception the plates of 2dF had been constructed from stainless steel. It is yet to be seen how this will affect the overall performance of 2dF with regard to fibre positioning.

The upcoming changes will take place from July 2016 when HERMES will be offline for a few months undergoing upgrades and maintenance. This will be followed in 2017 by upgrades to the fibre cables that feed HERMES and AAOmega. The HERMES upgrades aim to improve the optical elements to remove persistent cosmic-ray-like artefacts in the raw images, and to upgrade the cryostats to allow the cameras to hold vacuum for longer than is currently possible with the system. In theory none of the HERMES upgrades should substantially affect the throughput of the fibres and spectrograph system as they are all after the slitblock in the light path. However improvements to the cryostat will mean that the focus of the cameras can be held better, which would result in an improvement in the point spread function of the images.\\

We have presented relative fibre-to-fibre throughputs for the 2dF/HERMES system based upon analysis of the first two years of GALAH survey data. This has identified that there is an amount of variation between fibres and retractors in throughputs. There is also some amount of chromatic variation for certain regions of the plate. The cause of this effect is not currently understood.

There has been no obvious temporal degradation of the fibres over the course of the survey, which is very encouraging for the future use of this fibre cable.

We stress that while these variations in fibre throughput do not directly impact the scientific yield of HERMES or the GALAH survey, these results indicate that it is extremely important to continue monitoring the throughputs of the fibres of 2dF/HERMES as the GALAH survey progresses, increasing both the time baseline, but also the amount of data. Improvements in HERMES, the fibre cables and 2dF are all planned for the short and medium terms which could effect the relative throughputs of fibres.

\section*{Acknowledgements}
JS thanks Chris Lidman at the AAO for sending him down this path.

We thank the anonymous referee for their helpful comments that have improved this manuscript. We also thank Russel Cannon for his insights into the 2dF system.

The data in this paper were based on observations obtained at the Australian Astronomical Observatory as part of the GALAH survey (AAO Programs 2013B/13, 2014A/25, 2015A/19).

This research has been supported in part by the Australian Research Council (ARC) funding schemes (grant numbers DP1095368, DP120101815, DP120101237, DP120104562, FS110200035 and FL110100012). SLM acknowledges the support of the ARC through grant DE140100598. DMN was supported by the ARC grant FL110100012. DBZ acknowledges ARC funding from FS110200035. TZ acknowledges funding support from the Slovenian Research Agency and ESA.

The following software and programming languages made this research possible: \textsc{2dfdr}, the 2dF Data Reduction software \citep{2015ascl.soft05015A}; Python, in particular the packages Astropy, a community-developed core Python package for Astronomy \citep{TheAstropyCollaboration2013}, and pandas; Tool for OPerations on Catalogues And Tables (\textsc{topcat}) \citep{Taylor2005}.

This research was made possible through the use of the AAVSO Photometric All-Sky Survey (APASS), funded by the Robert Martin Ayers Sciences Fund. This publication makes use of data products from the Two Micron All Sky Survey, which is a joint project of the University of Massachusetts and the Infrared Processing and Analysis Center/California Institute of Technology, funded by the National Aeronautics and Space Administration and the National Science Foundation.



\bibliographystyle{mnras}

\begin{thebibliography}{}
\makeatletter
\relax
\def\mn@urlcharsother{\let\do\@makeother \do\$\do\&\do\#\do\^\do\_\do\%\do\~}
\def\mn@doi{\begingroup\mn@urlcharsother \@ifnextchar [ {\mn@doi@}
  {\mn@doi@[]}}
\def\mn@doi@[#1]#2{\def\@tempa{#1}\ifx\@tempa\@empty \href
  {http://dx.doi.org/#2} {doi:#2}\else \href {http://dx.doi.org/#2} {#1}\fi
  \endgroup}
\def\mn@eprint#1#2{\mn@eprint@#1:#2::\@nil}
\def\mn@eprint@arXiv#1{\href {http://arxiv.org/abs/#1} {{\tt arXiv:#1}}}
\def\mn@eprint@dblp#1{\href {http://dblp.uni-trier.de/rec/bibtex/#1.xml}
  {dblp:#1}}
\def\mn@eprint@#1:#2:#3:#4\@nil{\def\@tempa {#1}\def\@tempb {#2}\def\@tempc
  {#3}\ifx \@tempc \@empty \let \@tempc \@tempb \let \@tempb \@tempa \fi \ifx
  \@tempb \@empty \def\@tempb {arXiv}\fi \@ifundefined
  {mn@eprint@\@tempb}{\@tempb:\@tempc}{\expandafter \expandafter \csname
  mn@eprint@\@tempb\endcsname \expandafter{\@tempc}}}

\bibitem[\protect\citeauthoryear{{AAO Software Team}}{{AAO Software
  Team}}{2015}]{2015ascl.soft05015A}
{AAO Software Team} 2015, {2dfdr: Data reduction software}, Astrophysics Source
  Code Library

\bibitem[\protect\citeauthoryear{Akiyama et~al.,}{Akiyama
  et~al.}{2008}]{Akiyama2008}
Akiyama M.,  et~al., 2008, in Lemke E. A.-E.~D.,  ed.,  \procspie\ Vol. 7018,
  Advanced Optical and Mechanical Technologies in Telescopes and
  Instrumentation. Marseille, France, pp 70182V--70182V--12,
  \mn@doi{10.1117/12.788968}, \url
  {http://proceedings.spiedigitallibrary.org/proceeding.aspx?articleid=790889}

\bibitem[\protect\citeauthoryear{Allington-Smith et~al.,}{Allington-Smith
  et~al.}{2002}]{AllingtonSmith2002}
Allington-Smith J.,  et~al., 2002, \mn@doi [\pasp] {10.1086/341712}, 114, 892

\bibitem[\protect\citeauthoryear{Barden et~al.,}{Barden
  et~al.}{2010}]{Barden2010}
Barden S.~C.,  et~al., 2010, in McLean I.~S.,  Ramsay S.~K.,   Takami H.,  eds,
   \procspie\ Vol. 7735, Ground-based and Airborne Instrumentation for
  Astronomy III. San Diego, California, USA, pp 773509--773509--19,
  \mn@doi{10.1117/12.856103}, \url
  {http://proceedings.spiedigitallibrary.org/proceeding.aspx?articleid=750972}

\bibitem[\protect\citeauthoryear{Brzeski, Case  \& Gers}{Brzeski
  et~al.}{2011}]{Brzeski2011}
Brzeski J.,  Case S.,   Gers L.,  2011, in Hatheway A.~E.,  ed.,  \procspie\
  Vol. 8125, Optomechanics 2011: Innovations and Solutions. pp
  812504--812504--6, \mn@doi{10.1117/12.896389}, \url
  {http://proceedings.spiedigitallibrary.org/proceeding.aspx?articleid=1341547}

\bibitem[\protect\citeauthoryear{Brzeski, Gers, Smith  \& Staszak}{Brzeski
  et~al.}{2012}]{Brzeski2012}
Brzeski J.,  Gers L.,  Smith G.,   Staszak N.,  2012, in McLean I.~S.,  Ramsay
  S.~K.,   Takami H.,  eds,  \procspie\ Vol. 8446, Ground-based and Airborne
  Instrumentation for Astronomy IV. \procspie, Amsterdam, Netherlands, p.
  84464N, \mn@doi{10.1117/12.924635}, \url
  {http://proceedings.spiedigitallibrary.org/proceeding.aspx?doi=10.1117/12.924635}

\bibitem[\protect\citeauthoryear{Chromey \& Hasselbacher}{Chromey \&
  Hasselbacher}{1996}]{Chromey1996}
Chromey F.~R.,  Hasselbacher D.~a.,  1996, \mn@doi [\pasp] {10.1086/133817},
  108, 944

\bibitem[\protect\citeauthoryear{Colless et~al.,}{Colless
  et~al.}{2001}]{Colless2001}
Colless M.,  et~al., 2001, \mn@doi [\mnras] {10.1046/j.1365-8711.2001.04902.x},
  328, 1039

\bibitem[\protect\citeauthoryear{{De Silva} et~al.,}{{De Silva}
  et~al.}{2015}]{Silva2015}
{De Silva} G.~M.,  et~al., 2015, \mn@doi [\mnras] {10.1093/mnras/stv327}, 449,
  2604

\bibitem[\protect\citeauthoryear{Drinkwater et~al.,}{Drinkwater
  et~al.}{2010}]{Drinkwater2010}
Drinkwater M.~J.,  et~al., 2010, \mn@doi [\mnras]
  {10.1111/j.1365-2966.2009.15754.x}, 401, 1429

\bibitem[\protect\citeauthoryear{Driver et~al.,}{Driver
  et~al.}{2011}]{Driver2011}
Driver S.~P.,  et~al., 2011, \mn@doi [\mnras]
  {10.1111/j.1365-2966.2010.18188.x}, 413, 971

\bibitem[\protect\citeauthoryear{Fabricant et~al.,}{Fabricant
  et~al.}{2005}]{Fabricant2005}
Fabricant D.,  et~al., 2005, \mn@doi [\pasp] {10.1086/497385}, 117, 1411

\bibitem[\protect\citeauthoryear{Gilbert et~al.,}{Gilbert
  et~al.}{2012}]{Gilbert2012}
Gilbert J.,  et~al., 2012, in Navarro R.,  Cunningham C.~R.,   Prieto E.,  eds,
   \procspie\ Vol. 8450, Modern Technologies in Space- and Ground-based
  Telescopes and Instrumentation II. p. 84501A, \mn@doi{10.1117/12.924502},
  \url
  {http://proceedings.spiedigitallibrary.org/proceeding.aspx?doi=10.1117/12.924502}

\bibitem[\protect\citeauthoryear{Heijmans, Gers  \& Faught}{Heijmans
  et~al.}{2011}]{Heijmans2011}
Heijmans J. a.~C.,  Gers L.,   Faught B.,  2011, in Optical Design and
  Engineering IV. Marseille, France, p. 81671A, \mn@doi{10.1117/12.897273},
  \url
  {http://proceedings.spiedigitallibrary.org/proceeding.aspx?doi=10.1117/12.897273}

\bibitem[\protect\citeauthoryear{Henden \& Munari}{Henden \&
  Munari}{2014}]{Henden2014}
Henden A.,  Munari U.,  2014, Contributions of the Astronomical Observatory
  Skalnate Pleso, 43, 518

\bibitem[\protect\citeauthoryear{Henden, Levine, Terrell, Smith  \&
  Welch}{Henden et~al.}{2012}]{Henden2012}
Henden A.,  Levine S.,  Terrell D.,  Smith T.,   Welch D.,  2012, Journal of
  the American Association of Variable Star Observers (JAAVSO), 40, 430

\bibitem[\protect\citeauthoryear{Kimura et~al.,}{Kimura
  et~al.}{2010}]{Kimura2010}
Kimura M.,  et~al., 2010, \mn@doi [\pasj] {10.1093/pasj/62.5.1135}, 62, 1135

\bibitem[\protect\citeauthoryear{Lewis et~al.,}{Lewis et~al.}{2002}]{Lewis2002}
Lewis I.~J.,  et~al., 2002, \mn@doi [\mnras]
  {10.1046/j.1365-8711.2002.05333.x}, 333, 279

\bibitem[\protect\citeauthoryear{Miszalski, Shortridge, Saunders, Parker  \&
  Croom}{Miszalski et~al.}{2006}]{Miszalski2006}
Miszalski B.,  Shortridge K.,  Saunders W.,  Parker Q.~A.,   Croom S.~M.,
  2006, \mn@doi [\mnras] {10.1111/j.1365-2966.2006.10777.x}, 371, 1537

\bibitem[\protect\citeauthoryear{Munari et~al.,}{Munari
  et~al.}{2014}]{Munari2014}
Munari U.,  et~al., 2014, \mn@doi [\aj] {10.1088/0004-6256/148/5/81}, 148, 81

\bibitem[\protect\citeauthoryear{Newman}{Newman}{2002}]{Newman2002}
Newman P.~R.,  2002, \mn@doi [\pasp] {10.1086/341715}, 114, 918

\bibitem[\protect\citeauthoryear{Robitaille et~al.,}{Robitaille
  et~al.}{2013}]{TheAstropyCollaboration2013}
Robitaille T.~P.,  et~al., 2013, \mn@doi [\aap] {10.1051/0004-6361/201322068},
  558, A33

\bibitem[\protect\citeauthoryear{Sharp et~al.,}{Sharp et~al.}{2006}]{Sharp2006}
Sharp R.,  et~al., 2006, in McLean I.~S.,  Iye M.,  eds, Ground-based and
  Airborne Instrumentation for Astronomy. No. August in \procspie.
Orlando, Florida , USA, pp 62690G--62690G--13 (\mn@eprint {arXiv} {0606137}),
  \mn@doi{10.1117/12.671022}, \url
  {http://proceedings.spiedigitallibrary.org/proceeding.aspx?articleid=1288415}

\bibitem[\protect\citeauthoryear{Sharp, Brough  \& Cannon}{Sharp
  et~al.}{2013}]{Sharp2013}
Sharp R.,  Brough S.,   Cannon R.~D.,  2013, \mn@doi [\mnras]
  {10.1093/mnras/sts044}, 428, 447

\bibitem[\protect\citeauthoryear{Sheinis et~al.,}{Sheinis
  et~al.}{2014}]{Sheinis2014}
Sheinis A.,  et~al., 2014, in Ramsay S.~K.,  McLean I.~S.,   Takami H.,  eds,
  Ground-based and Airborne Instrumentation for Astronomy V. No.~June in
  \procspie\.
p. 91470Y, \mn@doi{10.1117/12.2055595}, \url
  {http://proceedings.spiedigitallibrary.org/proceeding.aspx?doi=10.1117/12.2055595}

\bibitem[\protect\citeauthoryear{Skrutskie et~al.,}{Skrutskie
  et~al.}{2006}]{Skrutskie2006}
Skrutskie M.~F.,  et~al., 2006, \mn@doi [\aj] {10.1086/498708}, 131, 1163

\bibitem[\protect\citeauthoryear{Taylor}{Taylor}{2005}]{Taylor2005}
Taylor M.,  2005, in Shopbell P.,  Britton M.,   Ebert R.,  eds,  Astronomical
  Society of the Pacific Conference Series Vol. 347, Astronomical Data Analysis
  Software and Systems XIV. p.~29

\bibitem[\protect\citeauthoryear{Wilson et~al.,}{Wilson
  et~al.}{2010}]{Wilson2010}
Wilson J.~C.,  et~al., 2010, in McLean I.~S.,  Ramsay S.~K.,   Takami H.,  eds,
   \procspie\ Vol. 7735, Ground-based and Airborne Instrumentation for
  Astronomy IV. Amsterdam, Netherlands, pp 77351C--77351C--14,
  \mn@doi{10.1117/12.856708}, \url
  {http://adsabs.harvard.edu/abs/2010SPIE.7735E..1CW
  http://proceedings.spiedigitallibrary.org/proceeding.aspx?articleid=750744}

\bibitem[\protect\citeauthoryear{Wyse \& Gilmore}{Wyse \&
  Gilmore}{1992}]{Wyse1992}
Wyse R. F.~G.,  Gilmore G.,  1992, \mn@doi [\mnras] {10.1093/mnras/257.1.1},
  257, 1

\bibitem[\protect\citeauthoryear{Zacharias, Finch, Girard, Henden, Bartlett,
  Monet  \& Zacharias}{Zacharias et~al.}{2012}]{Zacharias2013}
Zacharias N.,  Finch C.,  Girard T.,  Henden A.,  Bartlett J.,  Monet D.,
  Zacharias M.,  2012, \mn@doi [\aj] {10.1088/0004-6256/145/2/44}, 145, 14

\bibitem[\protect\citeauthoryear{Zheng, Gers  \& Heijmans}{Zheng
  et~al.}{2013}]{Zheng2013}
Zheng J.~R.,  Gers L.,   Heijmans J.,  2013, in F{\"{a}}hnle O.~W.,  Williamson
  R.,   Kim D.~W.,  eds,  \procspie\ Vol. 8838, Optical Manufacturing and
  Testing X. p. 88380G, \mn@doi{10.1117/12.2024493}, \url
  {http://proceedings.spiedigitallibrary.org/proceeding.aspx?doi=10.1117/12.2024493}

\makeatother
\end{thebibliography}

\appendix\section{Pivot-to-fibre mapping for 2dF/HERMES}
\begin{table*}
\caption{The mapping between the pivot number (order of fibres on the plate) and the fibre number (order of the fibres as seen by the slit and camera). The mapping is the same for Plate 0 and Plate 1, except for those entries with two fibre numbers, for which the first fibre number is for Plate 0 and the second for the Plate 1. Pivots marked with $^{\textrm{F}}$ are fiducial fibres. They have a fibre number but their light does not actually reach the slit.}
\label{table:fibretopivot}
\begin{tabular}{llllllllllllllllllll}
Pivot & Fibre &  & Pivot & Fibre &  & Pivot & Fibre &  & Pivot & Fibre &  & Pivot & Fibre &  & Pivot & Fibre &  & Pivot & Fibre \\
\cline{1-2} \cline{4-5} \cline{7-8} \cline{10-11} \cline{13-14} \cline{16-17} \cline{19-20} \\
1     & 10    &  & 61    & 70    &  & 121   & 130   &  & 181   & 190   &  & 241   & 250   &  & 301   & 310   &  & 361   & 370   \\
2     & 9     &  & 62    & 69    &  & 122   & 129   &  & 182   & 189   &  & 242   & 249   &  & 302   & 309   &  & 362   & 369   \\
3     & 8     &  & 63    & 68    &  & 123   & 128   &  & 183   & 188   &  & 243   & 248   &  & 303   & 308   &  & 363   & 368   \\
4     & 7     &  & 64    & 67    &  & 124   & 127   &  & 184   & 187   &  & 244   & 247   &  & 304   & 307   &  & 364   & 367   \\
5     & 6     &  & 65    & 66    &  & 125   & 126   &  & 185   & 186   &  & 245   & 246   &  & 305   & 306   &  & 365   & 366   \\
6     & 5     &  & 66    & 65    &  & 126   & 125   &  & 186   & 185   &  & 246   & 245   &  & 306   & 305   &  & 366   & 365   \\
7     & 4     &  & 67    & 64    &  & 127   & 124   &  & 187   & 184   &  & 247   & 244   &  & 307   & 304   &  & 367   & 364   \\
8     & 3     &  & 68    & 63    &  & 128   & 123   &  & 188   & 183   &  & 248   & 243   &  & 308   & 303   &  & 368   & 363   \\
9     & 2     &  & 69    & 62    &  & 129   & 122   &  & 189   & 182   &  & 249   & 242   &  & 309   & 302   &  & 369   & 362   \\
10    & 1     &  & 70    & 61    &  & 130   & 121   &  & 190   & 181   &  & 250$^{\textrm{F}}$   & 241   &  & 310   & 301   &  & 370   & 361   \\
11    & 20    &  & 71    & 80    &  & 131   & 140   &  & 191   & 200   &  & 251   & 260   &  & 311   & 320   &  & 371   & 380   \\
12    & 19    &  & 72    & 79    &  & 132   & 139   &  & 192   & 199   &  & 252   & 259   &  & 312   & 319   &  & 372   & 379   \\
13    & 18    &  & 73    & 78    &  & 133   & 138   &  & 193   & 198   &  & 253   & 258   &  & 313   & 318   &  & 373   & 378   \\
14    & 17    &  & 74    & 77    &  & 134   & 137   &  & 194   & 197   &  & 254   & 257   &  & 314   & 317   &  & 374   & 377   \\
15    & 16    &  & 75    & 76    &  & 135   & 136   &  & 195   & 196   &  & 255   & 256   &  & 315   & 315, 316   &  & 375   & 376   \\
16    & 15    &  & 76    & 75    &  & 136   & 135   &  & 196   & 195   &  & 256   & 255   &  & 316   & 316, 315   &  & 376   & 375   \\
17    & 14    &  & 77    & 74    &  & 137   & 134   &  & 197   & 194   &  & 257   & 254   &  & 317   & 314   &  & 377   & 374   \\
18    & 13    &  & 78    & 73    &  & 138   & 133   &  & 198   & 193   &  & 258   & 253   &  & 318   & 313   &  & 378   & 373   \\
19    & 12    &  & 79    & 72    &  & 139   & 132   &  & 199   & 192   &  & 259   & 252   &  & 319   & 312   &  & 379   & 372   \\
20    & 11    &  & 80    & 71    &  & 140   & 131   &  & 200$^{\textrm{F}}$   & 191   &  & 260   & 251   &  & 320   & 311   &  & 380   & 371   \\
21    & 30    &  & 81    & 90    &  & 141   & 150   &  & 201   & 210   &  & 261   & 270   &  & 321   & 330   &  & 381   & 390   \\
22    & 29    &  & 82    & 89    &  & 142   & 149   &  & 202   & 209   &  & 262   & 269   &  & 322   & 329   &  & 382   & 389   \\
23    & 28    &  & 83    & 88    &  & 143   & 148   &  & 203   & 208   &  & 263   & 268   &  & 323   & 328   &  & 383   & 388   \\
24    & 27    &  & 84    & 87    &  & 144   & 147   &  & 204   & 207   &  & 264   & 267   &  & 324   & 327   &  & 384   & 387   \\
25    & 26    &  & 85    & 86    &  & 145   & 146   &  & 205   & 206   &  & 265   & 266   &  & 325   & 326   &  & 385   & 386   \\
26    & 25    &  & 86    & 85    &  & 146   & 145   &  & 206   & 205   &  & 266   & 265   &  & 326   & 325   &  & 386   & 385   \\
27    & 24    &  & 87    & 84    &  & 147   & 144   &  & 207   & 204   &  & 267   & 264   &  & 327   & 324   &  & 387   & 384   \\
28    & 23    &  & 88    & 83    &  & 148   & 143   &  & 208   & 203   &  & 268   & 263   &  & 328   & 323   &  & 388   & 383   \\
29    & 22    &  & 89    & 82    &  & 149   & 142   &  & 209   & 202   &  & 269   & 262   &  & 329   & 322   &  & 389   & 382   \\
30    & 21    &  & 90    & 81    &  & 150$^{\textrm{F}}$   & 141   &  & 210   & 201   &  & 270   & 261   &  & 330   & 321   &  & 390   & 381   \\
31    & 40    &  & 91    & 100   &  & 151   & 160   &  & 211   & 220   &  & 271   & 280   &  & 331   & 340, 339   &  & 391   & 400   \\
32    & 39    &  & 92    & 99    &  & 152   & 159   &  & 212   & 219   &  & 272   & 279   &  & 332   & 339, 340  &  & 392   & 399   \\
33    & 38    &  & 93    & 98    &  & 153   & 158   &  & 213   & 218   &  & 273   & 278   &  & 333   & 338   &  & 393   & 398   \\
34    & 37    &  & 94    & 97    &  & 154   & 157   &  & 214   & 217   &  & 274   & 277   &  & 334   & 337   &  & 394   & 397   \\
35    & 36    &  & 95    & 96    &  & 155   & 156   &  & 215   & 216   &  & 275   & 276   &  & 335   & 336   &  & 395   & 396   \\
36    & 35    &  & 96    & 95    &  & 156   & 155   &  & 216   & 215   &  & 276   & 275   &  & 336   & 335   &  & 396   & 395   \\
37    & 34    &  & 97    & 94    &  & 157   & 154   &  & 217   & 214   &  & 277   & 274   &  & 337   & 334   &  & 397   & 394   \\
38    & 33    &  & 98    & 93    &  & 158   & 153   &  & 218   & 213   &  & 278   & 273   &  & 338   & 333   &  & 398   & 393   \\
39    & 32    &  & 99    & 92    &  & 159   & 152   &  & 219   & 212   &  & 279   & 272   &  & 339   & 332   &  & 399   & 392   \\
40    & 31    &  & 100$^{\textrm{F}}$   & 91    &  & 160   & 151   &  & 220   & 211   &  & 280   & 271   &  & 340   & 331   &  & 400   & 391   \\
41    & 50    &  & 101   & 110   &  & 161   & 170   &  & 221   & 230   &  & 281   & 290   &  & 341   & 350   &  &       &       \\
42    & 49    &  & 102   & 109   &  & 162   & 169   &  & 222   & 229   &  & 282   & 289   &  & 342   & 349   &  &       &       \\
43    & 48    &  & 103   & 108   &  & 163   & 168   &  & 223   & 228   &  & 283   & 288   &  & 343   & 348   &  &       &       \\
44    & 47    &  & 104   & 107   &  & 164   & 167   &  & 224   & 227   &  & 284   & 287   &  & 344   & 347   &  &       &       \\
45    & 46    &  & 105   & 106   &  & 165   & 166   &  & 225   & 226   &  & 285   & 286   &  & 345   & 346   &  &       &       \\
46    & 45    &  & 106   & 105   &  & 166   & 165   &  & 226   & 225   &  & 286   & 285   &  & 346   & 345   &  &       &       \\
47    & 44    &  & 107   & 104   &  & 167   & 164   &  & 227   & 224   &  & 287   & 284   &  & 347   & 344   &  &       &       \\
48    & 43    &  & 108   & 103   &  & 168   & 163   &  & 228   & 223   &  & 288   & 283   &  & 348   & 343   &  &       &       \\
49    & 42    &  & 109   & 102   &  & 169   & 162   &  & 229   & 222   &  & 289   & 282   &  & 349   & 342   &  &       &       \\
50$^{\textrm{F}}$    & 41    &  & 110   & 101   &  & 170   & 161   &  & 230   & 221   &  & 290   & 281   &  & 350$^{\textrm{F}}$   & 341   &  &       &       \\
51    & 60    &  & 111   & 120   &  & 171   & 180   &  & 231   & 240   &  & 291   & 300   &  & 351   & 360   &  &       &       \\
52    & 59    &  & 112   & 119   &  & 172   & 179   &  & 232   & 239   &  & 292   & 299   &  & 352   & 359   &  &       &       \\
53    & 58    &  & 113   & 118   &  & 173   & 178   &  & 233   & 238   &  & 293   & 298   &  & 353   & 358   &  &       &       \\
54    & 57    &  & 114   & 117   &  & 174   & 177   &  & 234   & 237   &  & 294   & 297   &  & 354   & 357   &  &       &       \\
55    & 56    &  & 115   & 116   &  & 175   & 176   &  & 235   & 236   &  & 295   & 296   &  & 355   & 355, 356   &  &       &       \\
56    & 55    &  & 116   & 115   &  & 176   & 175   &  & 236   & 235   &  & 296   & 295   &  & 356   & 356, 355   &  &       &       \\
57    & 54    &  & 117   & 114   &  & 177   & 174   &  & 237   & 234   &  & 297   & 293, 294   &  & 357   & 354   &  &       &       \\
58    & 53    &  & 118   & 113   &  & 178   & 173   &  & 238   & 233   &  & 298   & 292, 293   &  & 358   & 353   &  &       &       \\
59    & 52    &  & 119   & 112   &  & 179   & 172   &  & 239   & 232   &  & 299   & 291, 292  &  & 359   & 352   &  &       &       \\
60    & 51    &  & 120   & 111   &  & 180   & 171   &  & 240   & 231   &  & 300$^{\textrm{F}}$   & 294, 291   &  & 360   & 351   &  &       &   
\end{tabular}
\end{table*}


\bsp	
\label{lastpage}
\end{document}